\documentclass[preprint,authoryear,12pt]{elsarticle}

\usepackage{epsfig}

\usepackage{amssymb}

\usepackage[ps2pdf,%
a4paper=true,%
breaklinks=true,%
colorlinks=true,%
pdfauthor={Dudorov et al.},%
pdftitle={Theory of fossil magnetic field}%
]{hyperref}

\journal{Advances in Space Research}

\newcommand{\lti}[1]{\mbox{{\footnotesize #1}}}

\begin{document}

\begin{frontmatter}

\title{Theory of fossil magnetic field}

\author{Alexander E. Dudorov\corref{cor}}

\ead{dudorov@csu.ru}

\author{Sergey A. Khaibrakhmanov}
\address{Chelyabinsk state university, Bratiev Kashirinykh st., 129, Chelyabinsk, Russia, 454001}
\cortext[cor]{Corresponding author}
\ead{khaibrakhmanov@csu.ru}

\begin{abstract}

Theory of fossil magnetic field is based on the observations, analytical estimations and numerical simulations of magnetic flux evolution during star formation in the magnetized cores of molecular clouds. Basic goals, main features of the theory and manifestations of MHD effects in young stellar objects are discussed.

\end{abstract}

\begin{keyword}

young stars\sep accretion disks \sep magnetic fields \sep diffusion

\end{keyword}

\end{frontmatter}

\parindent=0.5 cm

\section{Introduction}

Hypothesis of fossil magnetic field was proposed by \cite{cowling45}, who showed that time scale of Ohmic dissipation of the magnetic field in the stars with masses $M \geq 1.5-2\,M_{\odot}$ exceeds time scale of their evolution. From this fact, Cowling concluded that stellar magnetic fields could be remnant (relic) of the magnetic field of protostellar clouds. 

Hypothesis of the fossil magnetic field \citep{spitzer58, mestel67} in its original form predicted that stellar magnetic fluxes exceed observational fluxes in $10^3-10^4$ times. From the other hand, estimations of \cite{ms56}, \cite{s68} showed that fossil magnetic flux can dissipate almost completely during protostellar clouds evolution.

Investigations performed since 1970 transformed hypothesis of fossil magnetism into the theory proving that, at least in young stars, magnetic field has fossil nature.

\cite{nt72} made first quantitative estimations of the magnetic field defreezing factor, determined as relation $\beta_f = B/B_f$, where $B_f$ is the frozen-in field, $B$ -- real field. 
Inclusion of ionization by cosmic rays and assumption about constancy of recombinations coefficients let them obtain any arbitrary small values of defreezing factor $\beta_f$.

\cite{dudorov77a} included mineral grains and thermal ionization of metals with low ionization potential (K, Na, Mg, Al, Ca) into consideration and estimated defreezing factor of the fossil magnetic field in collapsing interstellar clouds. He concluded that Ohmic diffusion and magnetic ambipolar diffusion are the main factors of fossil magnetic flux dissipation. Intensive dissipation of the magnetic field occurs in the density range $n=2\times 10^9-2\times 10^{11}\,\mbox{cm}^{-3}$, where $B\propto \rho^{2/5}$. In this density range, Ohmic diffusion and magnetic ambipolar diffusion reduce intensity of the magnetic field by factor $\beta_f=10^{-2}-10^{-4}$ comparing to the frozen-in field. \cite{ruz85} obtained similar estimation only for the potassium. These papers as well as paper by \cite{bs82} showed that defreezing factor of magnetic field in collapsing protostellar clouds is determined by their ionization and thermal structure, and also by degree of their non-uniformity and anisotropy.

\cite{nu86a, nu86b} carried out numerical estimations taking into account thermal processes and non-uniformity of the self-similar collapse. They estimated conditions of magnetic field defreezing assuming that typical time scale of Ohmic diffusion is equal to the free-fall time, $t_{od}=t_{ff}$. Obtained intensities of the magnetic field inside stars $B\simeq 1-10$ Gs are too small comparing to the magnetic field intensity given by the battery effect.

The theory of the fossil magnetic field has been further developed by \cite{dud81, dud82, dud87}. We present main result of these works in the following section. Main purpose of the fossil magnetic field theory is investigation of the magnetic field evolution of the protostellar clouds in process of its induction amplification and dissipation due to Ohmic diffusion and magnetic ambipolar diffusion. \cite{dud87} calculated not only surface ($1-100$ Gs), but also internal magnetic field $\sim 10^5-10^6$ Gs of young and main sequence stars.

Evolution of fossil magnetic field in process of protostar formation has been investigated by Mouschovias group \citep[see][and references therein]{dm01, km10}. They considered magnetic ambipolar diffusion as the main factor of fossil magnetic field evolution \citep{fm93}. Magnetic decoupling causes concentration of magnetic flux outside the regions of magnetic ambipolar diffusion \citep{tm05a, tm05b}. \cite{km10} investigated evolution of protostellar cloud core using numerical simulations in frame of multi-component 2D approach. They traced formation of the protostar. Characteristics of the protostar formed are close to observational, core density $n\simeq 10^{14}\,\mbox{cm}^{-3}$, mass $M\simeq 0.01\,M_{\odot}$, temperature $T=300$ K.

Applications of the theory of fossil magnetic field are discussed in many papers \citep[see][and references therein]{dudorov95, glag03, moss03, a08, br12}. \cite{dudorov90a} concluded that magnetic Cp stars are formed under the intensive action of cosmic rays. \cite{dudorov95} showed that convection in T Tauri and Ae/Be Herbig stars may lead to removal of fossil magnetic flux excess and to switching on the dynamo on the level of non-linear stabilization. Estimations and discussions of \cite{moss03} correspond to fully convective stars on the Hayashi track \citep{h62} and cannot be applied to the Ae/Be Herbig stars considered as the prototype of Cp stars. \cite{glag03} and \cite{a08} discussed observational data on magnetic fields and rotation of Cp stars and concluded that magnetic field of such stars may be the fossil magnetic field of pre-main-sequence stars.

The paper is organized as follows. In the section \ref{Sec:outine}, we review main statements and conclusions of the fossil magnetic field theory. In the section \ref{Sec:dynamo}, we discuss role of dynamo mechanisms in the magnetic field evolution during star formation. Main results of the 2D MHD numerical investigations of the magnetic protostellar clouds evolution are presented in the section \ref{Sec:2D}. In the section \ref{Sec:AD}, equations of the MHD model of accretion disks of young stars are formulated.  Then, we present results of calculations of the fossil magnetic field intensity and geometry in accretion disks of young stars. In the section \ref{Sec:Concl}, we outline main problems for the future investigations in frame of fossil magnetic field theory.

\section{Outline the theory of fossil magnetic field}
\label{Sec:outine}
Analysis of observational data on the magnetic field in the star forming regions \citep[see][]{dudorov90, dudorov91b, dudorov95} shows that there is correlation between the intensity, $B$, and  numerical density of clouds, $n$, for the star forming clouds,
 \begin{equation}
 B/B_0 = (n/n_0)^{k}. \label{eq:B} 
 \end{equation}
Parameters $B_0$ and $n_0$ are associated with cloud formation conditions. A large part of the observational data for regions  with densities $n\approx 1-10^{10}\, \mbox{cm}^{-3}$ is described by  formula (\ref{eq:B}) with parameters
 \begin{equation}
 B_0=5 \,\mu\mbox{Gs},\, n_0=50-200\, \mbox{cm}^{-3},\, k=1/2-1/3. \label{eq:par}
 \end{equation}
Optical and infrared polarimetry of dense molecular and protostellar clouds give very important confirmation and extension of Zeeman's observations \citep[see][]{vallee97, girart06, li09, chapman13}. Polarimetrical data show that the magnetic  field geometry is changed during the formation and evolution of interstellar clouds.

Magnetic field of the diffuse interstellar clouds is quite homogeneous with magnetic  strength  $\bf{B} \perp \bf{a}$ in $\approx 60\%$ of cases and $\bf{B} \parallel \bf{a}$ in $\approx 30\%$ of cases, where $\bf{a}$ is the direction of cloud elongation. In $\approx 10\%$ of cases, the magnetic fields of diffuse clouds have a tangled geometry. In more dense inhomogeneous cores of molecular clouds magnetic fields may have a quasi-radial, hourglass, twisted or pinched geometry.

The comparison of ``magnetic'' observational data for stars and for interstellar molecular clouds (and their cores) allows us to make the following conclusion. Contemporary star formation takes place mainly in the magnetized molecular clouds. Magnetic flux of young and some part of main sequence stars may be a relic of the magnetic flux of molecular or protostellar clouds. Therefore the basic goals of the theory consist in investigations of protostellar clouds and star formation in the sufficiently magnetized matter and in proof that the magnetic field of young stars may have the fossil origin. We discuss below the outline and base consequences of the theory of the fossil magnetic fields developed by author with collaborators during last 30 years.

Theory of  fossil  magnetic  field  is  based  on  the   numerical investigations  of  star  formation  in  protostellar  clouds with magnetic fields.  The main goals of the theory are  the  study  of evolution of magnetic flux in the processes of ambipolar and Ohmic diffusion,  interaction  with   rotation,   turbulence   and   MHD instabilities   that   may   develop  on  the  various  stages  of protostellar clouds collapse, protostar and star contraction. For these investigations we use the system of MGD (magneto-gas-dynamical) equations in the ``diffusional'' variables that contains the usual system of MGD equations and following additional equations  \citep[see][]{dudorov95},
\begin{equation} 
\frac{\partial x}{\partial t } + (\textbf{V}\cdot\nabla) x=\frac{S_p}
{\rho}-\frac{\nabla(x\rho\textbf{V}_{\lti{ad}})}{\rho},           \label{eq:x}
\end{equation}
\begin{equation}
\frac{\partial \textbf{V}_{\lti{ad}}}{\partial t}+ (\textbf {V}\cdot\nabla)
\textbf{V}_{\lti{ad}}=\frac{V_s^2}
{\rho_i}\frac{\nabla x}{x}+\frac{\textbf{F}_{em}}{x\rho}-\eta_{in}\rho\textbf{V}_{\lti{ad}},   \label{eq:vm}
\end{equation}
\begin{equation}
\frac{\partial \textbf{B}}{\partial t} =\mbox{rot}[(\textbf {V}+\textbf{V}_{\lti{ad}})\times \textbf
{B}]-\nabla\times(\nu_m\nabla\times\textbf{B}),\, \mbox{div}\textbf{B}=0, \label{eq:Bind}
\end{equation}
where  $ x=\rho_p/(\rho_p+\rho_n), \textbf{V}_{\lti{ad}}=\textbf{V}-\textbf{V}_p$  are  the  ionization degree and velocity of ambipolar diffusion,  $\rho_p$ and $\rho_n$  are densities  of  charge  and neutral components,  $\eta_{\lti{in}}$ -- coefficient of the momentum transfer between ions and neutrals, $\textbf{V}_p$ is the velocity of charge component, $S_p$ is the source function for the charges.  Other  quantities  are  used  in the usual astrophysical meaning. 
     
     The equations   (\ref{eq:x}--\ref{eq:Bind})   allow   us   to investigate  the  non stationary  and non equilibrium ionization and nonstationary magnetic  ambipolar  diffusion (MAD).
     
     We elaborated the one and half approximation for the numerical  solution of MGD system in the diffusional variables in the case of weak-magnetic  approach  \citep{dud81, dud82, dud87}.  The numerical simulations  are  carried  with  the  help  of  modified Lax-Vendroff  scheme.  We examined the thermal and ionization  history  in  detail.  We  investigated  ionization by cosmic rays, X-rays and radioactive  elements  taking into account   the radiative   recombinations   and recombinations on the grains. We considered the thermal ionization of trace  elements,  hydrogen and  helium  and  grains  evaporation. Therefore  we  can  study  very  carefully the ambipolar and Ohmic diffusion,  magnetic detachment and interaction of magnetic  field with rotation.
     
    The numerical calculations show, that the magnetic field is frozen in  gas during the isothermal collapse and it acquires with time a quasi-radial geometry besides the core. If the ionization state is determined  by  cosmic rays and radioactive elements,  MAD diminishes the magnetic field in  the  opaque  protostellar  core,  when  the central density $n_c\in [10^5n_0,  10^9n_0]$, where $n_0=10^4-10^5 \,\mbox{cm}^{-3}$ is  the  initial  density  of  protostellar  clouds.  The adiabatic heating of opaque core switches on the thermal evaporation of grains and thermal ionization of trace  elements  K,  Na,  Al  with lower ionization potentials \citep{dudorov77a}. Therefore in the regions with temperature $T\geq 4000-5000$ K the  magnetic field  will  be  immersed  into  the whole gas again.  The zone of  powerful MAD moves to the surface in process  of  protostars  and young  stars  evolution  coinciding  with  the  region  of minimal ionization degree.  The attenuation of  magnetic  field  comparing to the frozen-in  field  is  equal to $\approx 10^{-2}$ for the Sun-like stars on the birth line.
    
       The theory of fossil magnetic field has now a large number of applications \citep[see][]{dudorov90}.  In the frame of this theory we have studied the magnetic braking during star formation; formation of  protostellar  magnetospheres;  evolution  of internal magnetic field  of  normal  hot  main  sequence  stars;  ionizational   and diffusional  pumping  of the magnetic field of chemically peculiar stars;  escape of fossil magnetic field in the cool young star  as the  consequence  of  convective  destruction  of  regular  field, formation of magnetic flux tubes and their Ohmic decay and  rising to the surface; dynamics of magnetic flux tubes and attenuation of magnetic activity of young stars.
       
In frame of theory of fossil magnetic field following approximation for the intensity of the stellar surface magnetic field can be obtained
\begin{equation}
B_s\approx 200\left(\frac{\tau^2_{\kappa}}{T^2_0}\right)\left(\frac{M_0}{M_\odot}\right),  \label{eq:bsa}
\end{equation}
where grain  recombinations are taken into account, the range of cosmic rays is $R_{\lti{CR}}\approx 130\,\mbox{g}\, \mbox{cm}^{-2}$, $T_0\approx 10$ K and $M_0$  are  the  initial temperature and mass of protostellar cloud, $\tau_{\kappa}\approx 10\, \mbox{g}\,\mbox{cm}^{-2}$ is infrared optical depth \citep{dudorov90a}.

     Numerical calculations for the large sample  of various  factors,  determining  the  conditions  of  star  formation, allow us to obtain the following approximate formula for the ``defreezing'' factor  $\beta_f$
\begin{equation}
\beta_f\approx 10^{-2}\left(\frac{M}{M_\odot}\right)^{0.4-0.5},
M\approx 0.1-75 M_\odot.          \label{eq:bet}
\end{equation}

     The surface magnetic field (before the interaction  with convection),
\begin{equation}
B_s\approx \tau_{\lti{CR}}^{-2}\cdot Z_{\lti{RE}}\cdot(a/Y_{\lti{g}})\cdot\beta_f^{-1},
\label{eq:bsc}
\end{equation}
depends  on  the   ``optical''   depth   for cosmic   rays $\tau_{\lti{CR}}$, the abundance of  radioactive  elements  $Z_{RE}$, radius $a$ and abundance of grains $Y_{\lti{g}}$. For the contemporary values of this parameters
\begin{equation}
B_s\approx B_{s0}\cdot\left(\frac{M}{M_\odot}\right)^{0.25-0.35}.
\label{eq:bsc1}
\end{equation}
$B_{s0}\approx 1-100$ Gs for normal stars  and  $B_{s0}\le2000-3000$ Gs for the magnetic Cp stars.  The  strength  of  magnetic field is increased toward to the center of stars and in  the core has the values  $\approx  (1-10)\cdot10^6$ Gs  depending  on stellar mass.

\section{Dynamo in young stars}
\label{Sec:dynamo}
The investigation of convective  instability  of  plane 
parallel  layer  with magnetic field using the Boussinesq
approach for compressibility, and linear  theory
of  normal  perturbations  \citep{dudorov77b} allows us to formulate
the following ``convective theorem'': In young  stars with masses $M\le1.5\, M_\odot$  on  the  stage  of   gravitational contraction to main sequence,  dynamical  convection  may  develop independently of the magnetic field intensity if the magnetic energy $E_m< E_g$, where $E_g$  is  the modulus  of
gravitational energy, because of strong opacity of partial ionized hydrogen and helium and  infinite  power  of  gravitation source.

     Convective motions  are  turbulent,  because convective velocity $v_k\ge v_A$ Alfven's velocity, and Reynolds number $Re \gg Re_{cr}$, where $Re_{cr}$ is the critical Reynolds number. Turbulent convective motions in the rotating stars destroy the  regular  fossil  magnetic  field and generate a chaotic small-scale    turbulent    magnetic    field,  switch on  the dynamo mechanisms  after  decreasing  of  fossil  magnetic  field down to the strength lower than level of nonlinear dynamo stabilization. Small scale magnetic field is dissipated due to the buoyancy and magnetic ambipolar diffusion influencing the magnetic activity of young cool stars \citep{dudorov91a}.  
     
   Investigations of fossil magnetic  field  evolution  during  star formation show that activity of cool stars approaching the  main sequence must be supported by the dynamo mechanisms. The  precise time of switching of the fossil activity on  the  dynamo generated activity, $t\approx 0.3\, t_{KH}$, depends on the mass of stars and on the conditions of star formation. Near the switching time, the fossil magnetic  field  has  the  energy  $E_m\approx0.3\, E_{turb}$, which approximately equals the  level  of  nonlinear dynamo stabilization \citep[see][]{zeldovich87}
   
     The investigation of magnetic field evolution in  the  stars leads to statement of new problem: how can the  turbulent dynamo mechanisms develop under conditions, when the strength of the initial seed fossil magnetic field  is  close  to  the  level  of nonlinear stabilization or more? Intuitively we can say, that in such a situation the fossil magnetic field must be  supported at the level of nonlinear stabilization. This problem  requires detailed investigations. Our conclusions are  confirmed  by the scaling estimations \citep{dudorov95}.
     
Estimations of strength of toroidal and  poloidal  dynamo magnetic field components are made for the convective envelope model  of solar mass star situated near  to  the birth  line.  The dependence of basic parameters on the radial coordinate has great significance. The distribution of magnetic field inside the young star is close to the  power-low dependence on the  radius  with exponent $\approx 1.7$. The strength of poloidal field near the center  of  star  $B_p\approx  2\cdot10^6 $ Gs,  the  strength  of toroidal field is larger by order of magnitude.
     
     The estimations for sample of main sequence stars show that the strength of dynamo supported fossil magnetic field inside the young  stars  may  have  the  values  $\approx  (0.1-10)\cdot10^6$ Gs. Such a strength of toroidal magnetic field is quite enough for the formation of magnetic ropes and their rising  to  the  surface for Alfven time \citep{d89}.

\section{2D MHD simulations}
\label{Sec:2D}
We have created a new 2D numerical code in order to investigate various astrophysical MHD problems and to prove basic conclusions of the theory of fossil magnetic fields.  Our code is based on a simple TVD-scheme for the solution of MHD problems \citep{dudorov03a, dudorov03b}.  This scheme does not require calculating the eigenvectors of the hyperbolity matrix but only uses the maximal modules of the eigenvalues.  Scheme is monotonic and it has high resolution in the smoothing region of the solution.

         On the base of this scheme, the code ``Moon'' that can simulate MHD flows both in the 1D and 2D statement of the problem, was been developed.  It has a module to solve the Poisson equation for gravitational potential, permitting the computation of the gravitational collapse of protostellar clouds.  The Poisson equation is solved by the Douglas-Rachford ADI method.  The code is implemented using the programming language C++ and is completely object-oriented.  Our code was tested on many problems with known exact or approximate analytical solutions, demonstrating its capability for solving most MHD-problems.
         
    Numerical computations of the collapse of the magnetized protostellar cloud have confirmed earlier results in the frame of one and a half dimensional approximation \citep{dudorov90}.  The geometry of a magnetic field in a cloud with an initially small field becomes quasi-radial.  Magnetic fields lead to a flattening of collapsing clouds on the late stages of contraction.  When the initial magnetic field is strong, the collapse is switched to the quasi-static contraction. Such types of clouds must evolve on a diffusional time scale.
    
    Numerical simulations show that for sufficiently large values of the initial magnetic field, the degree of flattening is proportional to $\varepsilon_{m}^{-1/2}$, where $\varepsilon_{m}$ is the initial ratio of the magnetic energy to the gravitational one. For sufficiently strong magnetic fields, the central density at the free-fall time $t_{ff}$ is inversely proportional to $\varepsilon_{m}$ dependence. Let us stress that no rings are formed during early stage of protostellar collapse (before the formation of opaque core).
    
   We performed also the first simulations of the rotating magnetized clouds collapse with initial value $\varepsilon_{m}=0.2$ and initial relation of magnetic energy to kinetic energy of rotation, $\varepsilon_{m\omega}=1$.
The results of calculations show that flat fast rotating disk is formed around opaque core (protostar) in the free-fall time. Parameters of this disk are quite similar to parameters of accretion or protoplanetary disk for the system HH 4796 A.  The magnetic braking of cloud rotation (magnetic transport of angular momentum from cloud to interstellar medium) is inefficient comparing to magnetic redistribution of angular momentum inside the collapsing cloud. Last feature explains the possibility of formation of slow rotating protostar (and the star, after that) and fast rotating circumstellar disk.

    These estimations explain the observable rotation curve \citep{bouvier93} and rotational gap between T Tauri and  Ae/Be Herbig stars quite well.  The estimations show that the rotation  of young stars may be influenced by magnetic coupling of the star  with the circumstellar disk. The T Tauri stars with the mass  less than $\approx 2 M_{\odot}$, may have the open or dipole geometry  of the fossil magnetic field. The magnetic field of Herbig Ae/Be  stars with mass $2 M_{\odot} \leq M \leq 8 M_{\odot}$ may have the  quadrupole geometry of fossil magnetic field. The  precise values of rotational velocity also depend on such   parameters as surface magnetic field and   the accretion mass rate.
    
    These conclusions are consistent with the theory of fossil magnetic field that predict the formation of  magnetospheres on the stages of  protostars evolution.

\section{MHD model of accretion disks of young stars}
\label{Sec:AD}
We elaborated the kinematic MHD model of accretion disks of young stars \citep{dud14}. We consider stationary geometrically thin, optically thick accretion disk. Self-gravity of the disk is neglected comparing to the gravity of the star with mass $M$ situated at the origin. We assume that accretion disk is in hydrostatic equilibrium. Model allows us to carry out detailed investigation of intensity and geometry of fossil magnetic field in accretion disks of young stars.

Equations of the model are strictly derived from the basic system of the MHD equations \citep{landau_6} taking into account Ohmic and magnetic ambipolar diffusion \citep[e.g.,][]{dudorov95}. In the kinematic approximation we neglect the electromagnetic force in the equation of motion. 

In the adopted approximations, accretion disk structure is described by equations of the \cite{ss73} model. In addition to turbulent ``viscous'' heating of the accretion disk, we take into account contribution of the stellar radiation. Accretion disk of young stars are cold comparing to black holes accretion disks, so that dust grains are the main opacity agents. We use appropriate opacity evaluated according to \cite{semenov03}.

We show that frozen-in vertical magnetic field component is proportional to the surface density, $\Sigma$,
\begin{equation}
	B_z = B_{z0}\frac{\Sigma}{\Sigma_0},\label{Eq:BzFrozen}
\end{equation}
where $B_{z0}$ and $\Sigma_0$ correspond to the initial values of the magnetic field and density. We obtain estimation $B_{z0}(r=1\,\mbox{AU})=0.16\,\mbox{Gs}$ assuming that accretion disk is formed in the process of magnetostatic contraction from the protostellar cloud with density $10^{5}\,\mbox{cm}^{-3}$ and magnetic induction $10^{-5}$ Gs \citep{crutcher04}. This magnetic field intensity is in agreement with that obtained from measurements of the remnant magnetism of meteorites in the Solar system \citep{stacey76}.

In order to estimate intensity of the vertical magnetic field component taking into account magnetic ambipolar diffusion, we equate $B_z$ generation time $r/V_r$ ($V_r$ -- accretion speed) and magnetic ambipolar diffusion time $r/V_{\lti{ad}}$ that gives
\begin{equation}
	B_z = \left(4\pi\eta_{\lti{in}} x\rho^2 r\right)^{1/2}. \label{Eq:BzMAD}
\end{equation}

Accretion with speed $V_r$ and orbital rotation with speed $V_{\varphi}$ lead to generation of the radial, $B_r$,  and azimuthal, $B_{\varphi}$, magnetic field components from $B_z$. Diffusion of $B_r$ and $B_{\varphi}$ takes place in the $z$ direction mainly. For stationary geometrically thin disk, solutions for the radial and azimuthal components of the induction equation
\begin{eqnarray}
	B_r &=& -\frac{V_r z}{\eta}B_z,\label{Eq:BrSol}\\
	B_{\varphi} &=& -\frac{3}{2}\left(\frac{z}{r}\right)^2\frac{V_{\varphi} z}{\eta}B_z\label{Eq:BphiSol},
\end{eqnarray}
where total magnetic diffusion coefficient $\eta$ includes coefficients of the Ohmic and magnetic ambipolar diffusion.

Efficiency of the Ohmic and magnetic ambipolar diffusion depends on the ionization fraction. Our ionization model includes equation of the ionization balance that takes into account radiative recombinations and recombinations on the dust grain. Dust grains evaporation is taken into account. We consider ionization by cosmic rays, X-Rays and radioactive elements. In addition, we consider thermal ionization of ``mean'' metal, Hydrogen and Helium. ``Mean'' metal ionization potential and abundance logarithm are calculated as the averaged mean of corresponding parameters of Potassium, Sodium, Magnesium, Calcium and Aluminium \citep{dud87}.

\subsection{Illustrative results}
System of the our kinematic MHD model equations is non-linear system of algebraic equations.  It is solved by iterative methods. Here, we present illustrative results of the calculation of the ionization fraction and magnetic field components in the accretion disk of solar mass T Tauri star with mass $M=1\,M_{\odot}$, luminosity $L_{\star}=1\,L_{\odot}$, radius $R_{\lti{s}}=2\,R_{\odot}$ and intensity of the surface magnetic field $B_{\lti{s}}=2\,\mbox{kGs}$. Corresponding mass accretion rate $\dot{M}=10^{-8}\,M_{\odot}/\mbox{yr}$, turbulence parameter $\alpha=0.01$.  We adopt standard dust-to-gas mass fraction $Y_{\lti{g}}=0.01$, initial dust size $a_{\lti{d}}=0.1\,\mu$m. Typical parameters of cosmic rays and X-rays ionization rate are adopted \citep{spitzer68, casanova95}.  Ionization rate by stellar X-rays is calculated using approximation from \cite{bai09}.

\subsubsection{Ionization fraction}
\begin{figure}[htb!]
\begin{center}
\includegraphics*[width=0.95\textwidth]{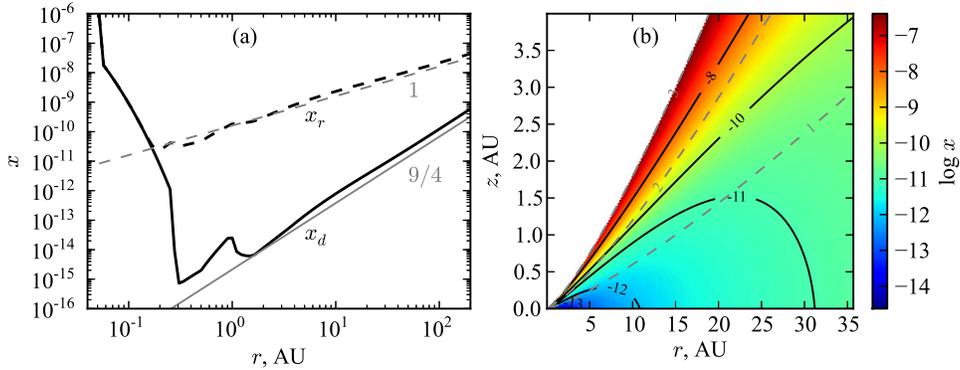}
\end{center}
\caption{Left: Dependence of midplane ionization fraction on $r$ (solid: with dust recombinations, dashed: with radiative recombinations), Right: ionization fraction distribution in the $r-z$ plane (filling and black contours; gray contours depict 1, 2 and 3 accretion disk scale heights). (A color version of this figure is available in the online journal.)}
\label{figure1}
\end{figure}

In the Fig(\ref{figure1}a), we plot radial profiles of the ionization fraction in the midplane of the accretion disk. Fig(\ref{figure1}b) shows distribution of the ionization fraction in the $r-z$ plane.

Ionization fraction has the minimum at $r_{\lti{min}}\sim 0.5$ AU. Thermal ionization operates at $r<r_{\lti{min}}$ and ionization fraction is large. Density decreases with distance, so cosmic rays and X-Rays ionize outer regions more effectively, and ionization fraction increases with distance at $r>r_{\lti{min}}$. Small peak at $r\sim 1$ AU is due to evaporation of ice dust grains.

In the case of the recombinations on the dust grains, ionization fraction falls down to extremely low value, $x_{\lti{min}}\sim 10^{-15}$ at $r_{\lti{min}}$. In the absence of dust, ionization fraction is larger be several orders of magnitude, so that $x_{\lti{min}}\sim 10^{-11}$. Fig(\ref{figure1}b) demonstrates layered stricture of the accretion disk. Cosmic rays and X-rays ionize effectively surface layers of the accretion disk, so that $x\sim 10^{-6}$ at heights $z\sim 3\,H$.

\subsubsection{Fossil magnetic field geometry}
\begin{figure}[htb!]
\begin{center}
\includegraphics*[width=8.8cm]{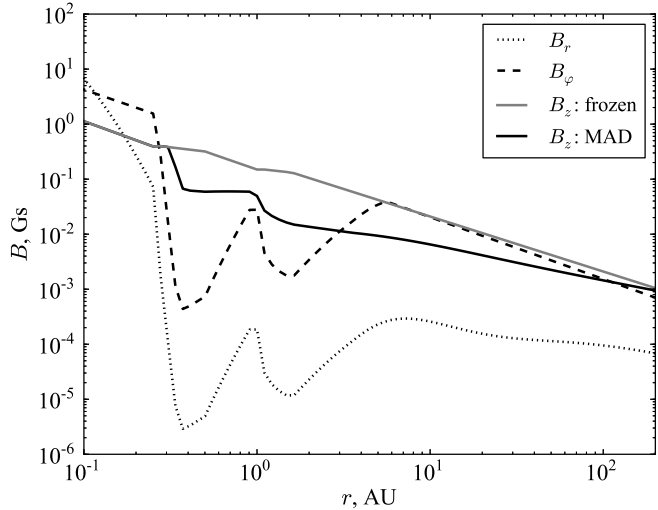}
\end{center}
\caption{Dependence of the magnetic field components evaluated at height $z=H$ on radial distance. Dots: absolute value of $B_r$, dashes: absolute value of $B_{\varphi}$, gray solid line: frozen-in $B_z$, black solid line: $B_z$ calculated taking into account magnetic ambipolar diffusion.}
\label{figure2}
\end{figure}

In the fig(\ref{figure2}), we plot radial profiles of the absolute values of the magnetic field components evaluated at the height $z=H$ above the midplane. In the calculation, recombinations on the dust grains are taken into account. For comparison, figure shows both frozen-in field $B_z$ and that calculated taking into account magnetic ambipolar diffusion. 

Our calculations show that there is three regions with the different geometry of the fossil magnetic field in the accretion disk, in the case of efficient recombinations on the dust grains. Diffusion of the magnetic field is inefficient close to the accretion disk inner edge $r\lesssim 0.1$ AU, where thermal ionization operates. Magnetic field is frozen-in here and it has quasi-azimuthal geometry, $B_{\varphi}>B_z,\,B_r$. Generation of the dynamically strong magnetic field is possible here. Intensity of the $B_{\varphi}$ is comparable with the intensity of the stellar magnetic field at the accretion disk inner edge.

In the region of the smallest ionization fraction, so-called ``dead'' zone \citep{gammie96}, $r\gtrsim 0.1$ AU and $r\lesssim 10-20$ AU, efficient magnetic diffusion prevents generation of $B_r$ and $B_{\varphi}$. Magnetic field remains quasi-poloidal in this region, $B_z > B_{\varphi} \gg B_r$, since times of the Ohmic and magnetic ambipolar diffusion are small compared to dynamical time here. Intensity of the $B_z$ at 3 AU is $\sim$0.01 Gs which is order of magnitude smaller than intensity of the frozen-in magnetic field.

Our calculation show that magnetic ambipolar diffusion of $B_{\varphi}$ is inefficient in the outer regions of the accretion disk, $r>10-20$ AU. Azimuthal magnetic field component is frozen-in here and its intensity is comparable with the intensity of the $B_z$. Buoyancy limits $B_{\varphi}$ in this region. Magnetic ambipolar diffusion prevents generation of the radial magnetic field component. Thus, fossil magnetic field has quasi-azimuthal geometry in considered conditions. Big grains or enhanced ionization rates are needed for magnetic field to be quasi-radial in the outer regions of the accretion disks \citep{dud14}.

In the absence of the dust, ionization fraction $x\gtrsim 10^{-11}$ and magnetic diffusion is inefficient in the whole accretion disk. Magnetic field is frozen-in, under such circumstances. Magnetic field is quasi-azimuthal in the inner regions of the accretion disk, $r<1$ AU, and quasi-radial in its outer regions.

\section{Conclusion}
\label{Sec:Concl}
Over past 40 years, the modern theory of star formation in magnetized molecular clouds has been successfully developed. Nevertheless, direct numerical simulation of the star formation in 2D and 3D taking into account all physical effects still remains a big challenge. The most detailed investigations were carried out by our group and by Mouschovias group. But, it is hard to say, when this problem will be solved completely.

Dynamical influence of the magnetic field is the missing ingredient in large part of theoretical studies of star formation. Kinematic approximation does not allow to investigate self-consistently evolution of stars and their accretion disks. This is the problem for future investigations.

In the last time, interest of researchers of chemically peculiar stars in the theory of fossil magnetic field is growing. The important task is to connect modern observations with the predictions of the fossil magnetic field theory as it was done by \cite{dudorov90a}. Recently, several researchers \citep[e.g.][]{moss03, br12} rediscover some statements of the fossil magnetic field theory.

\end{document}